\documentclass[%
reprint,
 amsmath,
 amssymb,
floatfix,
]{revtex4-2}

\usepackage{graphicx}
\usepackage{dcolumn}
\usepackage{bm}
\usepackage[ruled,vlined,linesnumbered]{algorithm2e}

\begin{document}

\preprint{APS/123-QED}

\title{Hybrid Logical-Physical Qubit Interaction as a Post Selection Oracle}

\author{Nadav Carmel}%
\email{nadav.carmel1@huji.mail.ac.il}
\author{Nadav Katz}%
\email{nadav.katz@mail.huji.ac.il}
\affiliation{%
 The Racah Institute of Physics,\\
 The Hebrew University of Jerusalem, Givat Ram, Jerusalem 9190410, Israel\\
}%

\date{\today}

\begin{abstract}
We demonstrate a property of the quantum 5-qubit stabilizer code that enables the interaction between qubits of different logical layers, and conduct a full density-matrix simulation of an interaction between a logical and a physical qubit. We use the logical qubit as an ancilla and find under which circumstances it gives an advantage over the bare physical ancilla approach, changing the circuit depth and noise level with decoherence processes at play. We use it as a post selection oracle for quantum phase estimation to detect errors propagating from the sensor qubit. Finally, we use our simulation to give noise thresholds both for computation and for sensing a signal using quantum phase estimation that are well within the capabilities of today's hardware.
\end{abstract}

\maketitle


In the field of quantum metrology, we are most interested in finding ways to recover the Heisenberg-limit scaling, promising that the error in estimating our observable scales as one over the number of measurements, probing time or number of probes \cite{zhou2020saturating}. Recent results indicate that this limit cannot be recovered in the presence of general Markovian noise if the Hamiltonian lies in the span of the noise operators \cite{zhou2018achieving, layden2018spatial,rojkov2022bias,chen2021effects,matsuzaki2017magnetic}. Much effort has been made to recover the Heisenberg-limit scaling using quantum error correction \cite{kessler2014quantum, dur2014improved, ma2021adaptive, zhou2020saturating, reiter2017dissipative, herrera2015quantum, arrad2014increasing, unden2016quantum,kapourniotis2019fault,zhuang2020distributed,layden2018spatial,rojkov2022bias}. All these efforts focus on sequential quantum metrology, encoding the sensor as a logical qubit and using sophisticated methods to correct the errors while not correcting the signal itself. 

Degen \textit{et al.} \cite{degen2017quantum} quantitatively define the Dynamic Range of a non-entangled quantum sensor, and finds it scales as the square root of the measurement time due to shot noise. Algorithmic quantum sensing is crucial in extending the dynamic range of quantum sensors by assigning appropriate weights to different quantum measurements, thereby approaching the Heisenberg limit. Recently the use of algorithmic quantum sensing has caught the attention of the community \cite{corcoles2021exploiting, kapourniotis2019fault, paesani2017experimental, svore2013faster}. While Quantum Phase Estimation (QPE) is a prominent algorithmic sensing protocol it has various applications in other areas of study \cite{o2019quantum, cruz2020optimizing, santagati2018witnessing, daskin2014universal, tilly2022variational}, the most famous one being Shor's algorithm. Thus extensive research has investigated the performance of QPE under noise \cite{chapeau2020fourier, dobvsivcek2007arbitrary, garcia2008quantum, o2021error}. Achieving the Heisenberg-limit scaling requires an unbiased probability distribution \cite{degen2017quantum, chen2021effects}. However, decoherence introduces bias to the probability distribution obtained from QPE, hindering the attainment of the Heisenberg limit. In this letter, we propose a method to mitigate bias, thereby approaching the limit.

In algorithmic quantum sensing protocols, the presence of ancillas introduces a new realm to study wherein error correction or error detection is performed on the ancillas rather than the sensor itself. Making only the ancilla logical while letting the sensor remain a physical sensor, enforces the need of hybrid logical-physical interaction.
While fault-tolerant \cite{aharonov1997fault} hybrid interaction could be pursued using methods such as flag fault-tolerance \cite{chao2018fault, chao2020flag, reichardt2020fault, debroy2020extended}, for sensing purposes, a limited number of successful runs is sufficient. Applying post selection on the sensor has proved to be a vital tool for any experimental implementation of sensing \cite{varbanov2020leakage,kwon2023efficacy,o2021error,matsuzaki2017magnetic,yamamoto2022error,arvidsson2020quantum}. It has been shown that the error in verified phase estimation \cite{o2021error}, an error mitigation technique based on what we call sensor post-selection (SPS), scales as the squared probability for a single gate error $p^2$.

In this letter, we take advantage of error propagation from the sensor qubit to the logical ancilla qubit. We get rid of a larger portion of the noise by encoding some of it on the redundant degrees of freedom of the Hilbert space of the logical ancilla, applying only error detection and post-selecting the results. We use the simple 5-qubit code \cite{knill2001benchmarking,knill1997theory,laflamme1996perfect} which has an interesting attribute: all errors with a weight smaller or equal to 2 cause a non-trivial syndrome. Thus if the probability of error in one ancilla in the whole algorithm is $p$, then the probability of error after logically post-selecting (LPS) is proportional to $p^3$. See \cite{Supplemental} for the 5-qubit-code stabilizers and a syndrome-cause table.

\begin{figure*}
  \centering
  \includegraphics[width=1.0\textwidth]{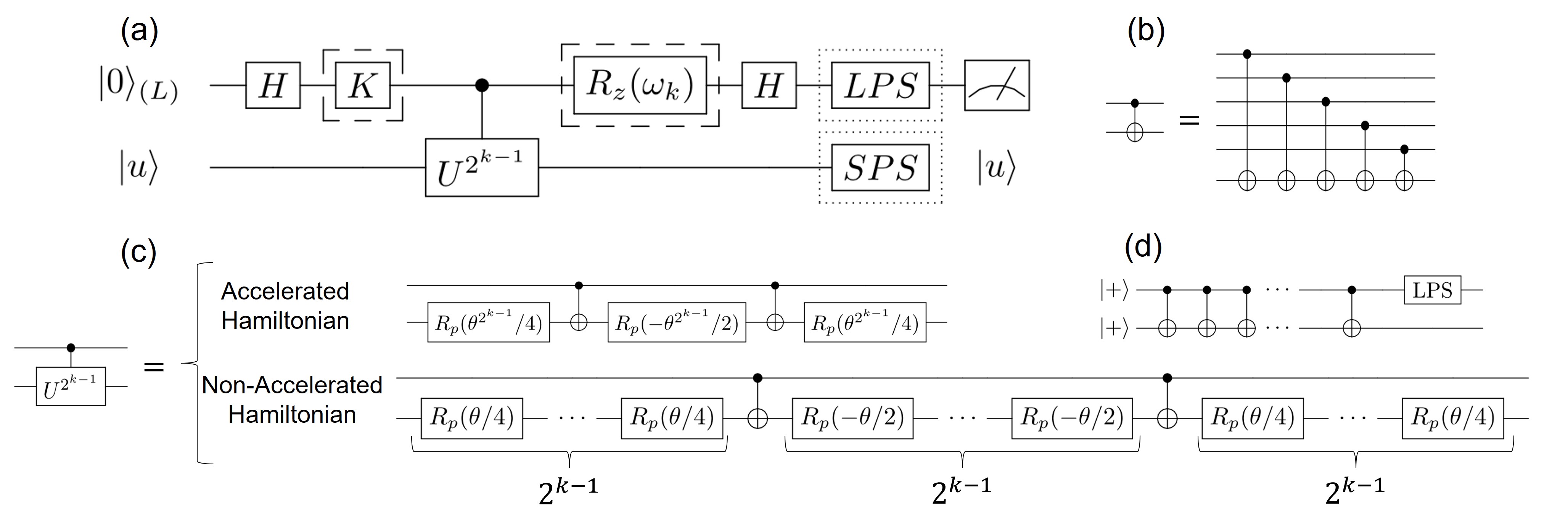}
  \caption{Circuits used in this work. (a) Iterative versions of QPE. The dashed $K$ gate appears only in Kitaev's iterative version, where $K$ can be $K=S$ and $K=I$. The dashed $R_z$ gate appears only in Iterative Phase Estimation Algorithm (IPEA), where the feedback angle depends on the previously measured bits through $\omega_k=-2\pi(0.0x_{k+1} x_{k+2} \cdots x_m)$, and $\omega _m = 0$. The dotted LPS (Logical Post Selection) and SPS (Sensor Post Selection) gates appear in circuits as described in the main text and other figures. (b) Logically controlled CNOT gate, with the first 5 qubits acting as a logical qubit and the sixth qubit as the target qubit, encoded one logical layer lower then the control. (c) The controlled operation in this work is a single qubit rotation gate where $p=x$ or $p=z$. In the figure are implementations of accelerated and non-accelerated controlled signal Hamiltonians. (d) The circuit used for exploring logical-physical interaction length, Fig. \ref{fig:computing_united} (b).}
  \label{fig:circuits}
\end{figure*}

Since any multi-qubit gate can be decomposed into single qubit gates and CNOT gates \cite{nielsen2002quantum}, our focus should be understanding how to implement the CNOT gate between logical and physical qubits, as control and target respectively. Some quantum error correction codes (QECC) have a useful parity attribute: The logical states, $|0\rangle_L$ and $|1\rangle_L$, are made up of a sum of quantum states with an even or odd number of $1$'s, respectively. One such code is the 5-qubit code \cite{laflamme1996perfect}, with basis states defined in \cite{Supplemental}. 
In the case of one logical layer, this attribute allows us to implement the CNOT gate in a semi-transversal manner, as in Fig. \ref{fig:circuits} (b). This can be generalized trivially for any number of logical layers, provided that the quantum code used for each layer has this attribute.
See \cite{Supplemental} for the logical gates  necessary for the QPE algorithm using the 5-qubit code.

Throughout the letter a fidelity between a 6-qubit-state and a 2-qubit-state has been calculated. This has been done by taking the 6-qubit-state's density matrix $\rho$ and calculating the effective reduced density matrix $\rho'$:
\begin{equation} \label{eq:reduced}
 \rho' = \begin{pmatrix}
\langle0_L0|\rho|0_L0\rangle & \langle0_L0|\rho|0_L1\rangle & \langle0_L0|\rho|1_L0\rangle & \langle0_L0|\rho|1_L1\rangle\\
\langle0_L1|\rho|0_L0\rangle & \langle0_L1|\rho|0_L1\rangle & \langle0_L1|\rho|1_L0\rangle & \langle0_L1|\rho|1_L1\rangle\\
\langle1_L0|\rho|0_L0\rangle & \langle1_L0|\rho|0_L1\rangle & \langle1_L0|\rho|1_L0\rangle & \langle1_L0|\rho|1_L1\rangle\\
\langle1_L1|\rho|0_L0\rangle & \langle1_L1|\rho|0_L1\rangle & \langle1_L1|\rho|1_L0\rangle & \langle1_L1|\rho|1_L1\rangle\\
\end{pmatrix}
\end{equation}
Note that this is not necessarily a pure state. Performing LPS in our simulation is done by projecting the state onto the code, forcing each stabilizer to measure '0' \cite{Supplemental}. In addition, when we perform error correction and the resulting state is not within the code, the reduction operation of Eq. \ref{eq:reduced} is not trace preserving. Due to the projective nature of these operations, to calculate the fidelity or distance between two states, where one of them is not a pure density matrix, we save the trace of each one and normalize them before hand. We define the lost information \cite{Supplemental} to be the fraction of information lost due to post selection, $l_i=1-Tr(\rho')$ where $\rho'$ is the reduced 2-qubit density matrix defined in Eq.\ref{eq:reduced}.
We quantify the noise by worst-case single gate fidelity, calculated by putting a qubit in it's most susceptible state to the applied noise - for example, the state $|+\rangle$ for dephasing and $|1\rangle$ for amplitude damping.

\begin{figure*}
  \centering
  \includegraphics[width=1.0\textwidth]{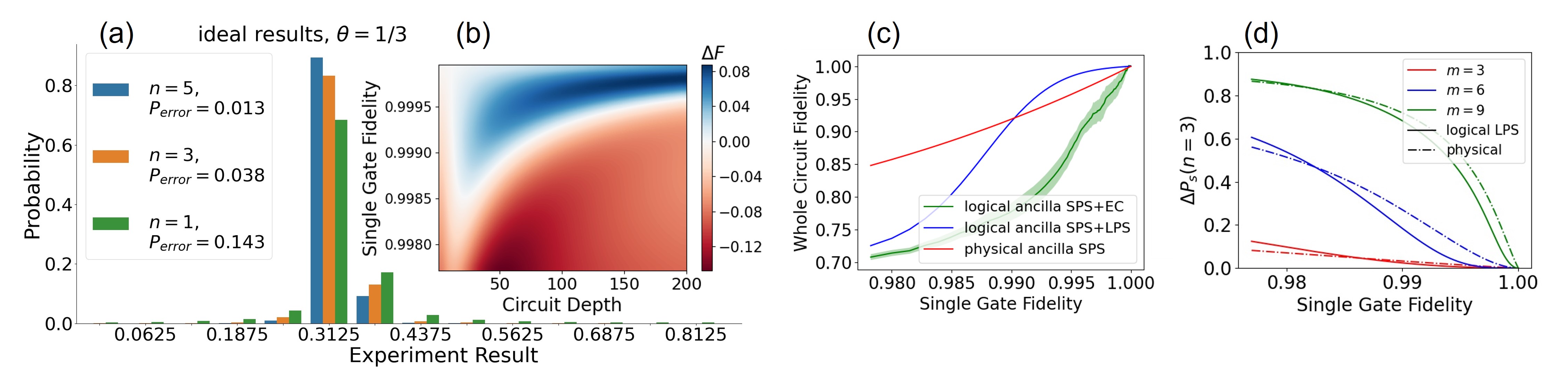}
  \caption{Demonstrating the general properties of LPS gadget under different scenarios. (a) Estimating $\theta = 1/3$ up to 4 binary digits. The resulting probability distributions are obtained from IPEA by measuring each digit $n\in \{1,3,5\}$ times and taking a majority vote. (b) Mapping the thresholds of logical and physical control for noise level and circuit depth, simulating the circuit in Fig. \ref{fig:circuits}d. Here all qubits are noisy (susceptible only to dephasing) in all stages of the circuit, including syndrome extraction. The white line is the location of the threshold, and blue/red areas are the fidelity difference at the end of the circuit $\Delta F = F_{logical} - F_{physical}$. A non-monotonic behaviour of the fidelity with circuit depth is observed. (c) Fidelity after an iteration of Kitaev's approach with accelerated sensing Hamiltonian averaged over 10 angles, with dephasing ancillas and a sensor in the ground state measuring $R_z(\theta)$. The red line represents physical ancilla with SPS, the blue line logical ancilla with SPS and LPS while the green line is logical ancilla with error correction right before the measurement. These graphs are for $K=I$ and similar graphs are available for $K=S$ in \cite{Supplemental}. It is clear that there is a threshold in which using LPS outputs a state closer to the ideal, even when the noise resides in the ancillas alone. (d) Difference between the ideal success probability of IPEA and a noisy implementation, both for physical and logical ancillas, for $n=3$. Here the ancillas are perfect and the sensor dephases from the $|+\rangle$ eigenstate of $R_x(\theta)$. The probabilities are extracted by summing the two binary results closest to the real measured angle, $2\pi/\sqrt{3}$.}
  \label{fig:computing_united}
\end{figure*}

Quantum phase estimation encompasses a family of algorithms for estimating the unknown phase, denoted as $\phi$, associated with the eigenvector $|u\rangle$ of a unitary operator $U$, having the eigenvalue $e^{2\pi i \phi}$. This estimation is achieved by employing black boxes capable of preparing the state $|u\rangle$ and performing controlled-$U^{2^{j}}$ operations, where $j$ is a positive integer. These controlled operations can be either accelerated or non-accelerated, where accelerated Hamiltonians have been extensively studied in the context of algorithmic complexity theory and super-resolution \cite{atia2017fast, aharonov2002measuring}. Acceleration can be achieved through specialized techniques like angle-dependent magnetic field application or general methods like QAQC \cite{khatri2019quantum}  and VFF \cite{cirstoiu2020variational}.

\begin{figure*}
  \centering
  \includegraphics[width=1.0\textwidth]{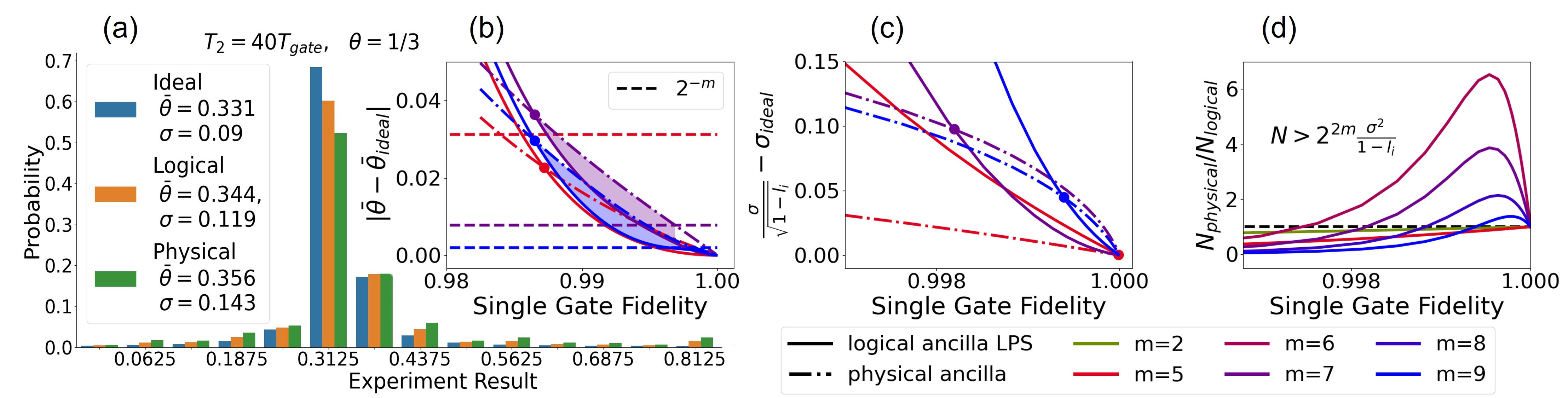}  \caption{Demonstrating LPS as a post selection oracle for IPEA (Fig.\ref{fig:circuits}a) in the context of quantum metrology. b-d are obtained by measuring $R_x(\frac{2\pi}{\sqrt{3}})$ on the $|+\rangle$ eigenstate, with a dephasing sensor and perfect ancillas. (a) Estimating $\theta = 1/3$ up to 4 binary digits. The resulting probability distributions are obtained from noisy IPEA algorithms as indicated in the figure. It is evident that LPS makes the probability distribution less biased, thus approaching the ideal and the Heisenberg limit. (b) Measuring the mean of the circular probability distribution of results for a number of desired precisions $m$. It is easy to see that there exists parameter regimes of deep circuits and approximately a minimum of 0.986 worst case single gate fidelity such that the logical control gives a better estimate of the mean. The shaded regions indicate on parameter regimes where it is beneficial to use LPS accuracy-wise. Lost information of the threshold is around $99\%$. (c) Statistical error in estimating the phase. (d) Ratio of the assessed minimum number of trials needed to reach digital accuracy by Eq.\ref{eq:Nmin}. It is evident from (c,d) that up from approximately 0.997 worst case single gate fidelity we start observing an improvement of results in the sense of a lower sensing time, while lost information of the threshold is around $75\%$.}
  \label{fig:sensing_united}
\end{figure*}

The algorithm employs two quantum registers: one for the measured operator $U$ and another for ancilla qubits needed for computation \cite{kitaev1995quantum}. In this study, we consider two iterative versions of the algorithm depicted in Fig. 1(a) (dashed line alternatives), which utilize a single ancilla qubit for phase estimation. These versions are particularly valuable for Noisy Intermediate Scale Quantum (NISQ) computers, where simultaneous utilization of multiple qubits is limited. We utilize Kitaev's approach \cite{Supplemental} with accelerated Hamiltonians, which is essentially equivalent to the shortest application of IPEA \cite{Supplemental}  in terms of fidelity. This approach allows us to average over measured phases and evaluate the performance of LPS (Fig. \ref{fig:computing_united} (b,c)). For unaccelerated Hamiltonians, we simulate the case using IPEA (Figs. \ref{fig:computing_united} (a,d),  \ref{fig:sensing_united} (a-d)).

Each QPE application yields a probability distribution of all possible measurement results, derived from the probabilities of correctly measuring each digit during each iteration or ancilla qubit. Multiple measurements of each digit can be performed, followed by a majority vote, to enhance the probability of accurate measurements, as illustrated in Fig.\ref{fig:computing_united} (a). Decoherence introduces bias to this histogram, altering the mean and standard deviation, as illustrated in Fig.\ref{fig:sensing_united}.

Quantum phase estimation finds utility in both quantum computation and quantum metrology, with different requirements in each. In quantum computation, the focus is on the probability of success after a single algorithm application. In quantum metrology, the primary interest lies in measuring a value and its associated error with the Heisenberg scaling.
In this study, we first assess the performance of the LPS gadget by measuring fidelity at the circuit's end and estimating the probability of a single-shot success in quantum phase estimation, disregarding any lost information (Fig.\ref{fig:computing_united}). Subsequently, we evaluate the algorithm's performance in the context of algorithmic sensing, accounting for lost information, as demonstrated in Fig.\ref{fig:sensing_united}.

Our first result is on the general behaviour of logical-physical interaction and LPS. As mentioned earlier, multi-qubit interactions can be simplified to single qubit gates along with CNOT or CZ entangling gates. A crucial question arises regarding the threshold value at which logical control surpasses physical control, and its dependence on the number of entangling gates. To address this question, we consider infinite $T_1$ and varying $T_2$ values for all qubits. Additionally, we vary the number of CNOT gates within the range of [1, 200), as depicted in Fig.\ref{fig:circuits} (d) (simulated circuit). We start with the initial eigenstate $|++\rangle$ of CNOT (the first qubit is the ancilla and the second is the sensor), which is highly susceptible to $T_2$ noise. The ancilla qubit is tested both as a logical 5-qubit or a bare physical qubit. After applying $N_g$ CNOT gates, we subject the system to noisy LPS. We calculate and save the fidelity between the output state and $|++\rangle$ using Eq.\ref{eq:reduced},  subsequently generating a color map illustrating the fidelity difference between logical and physical control (Fig.\ref{fig:computing_united} (b)). The number of entangling gates serves as one axis, while the worst-case single gate fidelity serves as the other. It should be noted that we employ noisy syndrome extraction, resulting in a constant overhead of approximately 20 gates in circuit depth. As depicted, there exists a range of circuit depths and gate fidelities wherein logical control outperforms physical control. The specific values obtained in our work are contingent on simulation parameters and are expected to vary with different quantum hardware. Nonetheless, the overall structure of the dependence is expected to stay similar.

Our second result is on the fidelity at the end of a QPE iteration. Algorithmic quantum sensing utilizing the QPE algorithm is particularly effective when accelerated Hamiltonians are assumed, allowing for high-fidelity application of high powers of the time evolution operator. In this letter, we consider accelerated Hamiltonians implemented as shown in Fig.\ref{fig:circuits} (c), simulating one iteration of Kitaev's approach (Fig.\ref{fig:circuits}(a)) with logical post-selection (LPS) and sensor post-selection (SPS). We calculate the fidelity, after post-selection, between the noisy implementation and the two-qubit ideal state immediately prior to measurement, thus assuming a perfect measurement.

The results, including the fidelity of the error-corrected state, are presented in Fig.\ref{fig:computing_united} (c). Notably, the error-corrected state exhibits stochastic behaviour, which is averaged in the figure due to its inherent randomness - a mistake in syndrome extraction leads to the application of a faulty correction operator, and it should be noted that we incorporate only one round of error correction in the circuit. Further details on the improved scaling of the error probability and complementary information can be found in \cite{Supplemental}.

Counter-intuitively, we observe a threshold at which logical control surpasses physical control. This is unexpected since only the ancillas are subject to noise. Apparently, there are scenarios in which employing five noisy ancillas outperforms using only one noisy ancilla. Strikingly, the obtained thresholds fall within the capabilities of today’s state-of-the-art technology! \cite{kandala2021demonstration}.

As mentioned earlier, it is sometimes advantageous to measure each digit multiple times and employ majority voting to enhance the success probability. In the NISQ era, we assume utilizing the deepest possible circuit for accurate phase measurement, where at most two possible results contribute to the algorithm's success probability (Fig.\ref{fig:computing_united}(d)). For an extended discussion on the success probability please refer to \cite{Supplemental}. It is evident that by acting as a post selecting oracle the LPS gadget increased the probability of success after a single run for a wide range of parameters.

In quantum metrology, the precision and statistical error of measurements are of utmost importance. Our previous findings demonstrate that the LPS gadget in deep quantum circuits has the potential to mitigate bias in the resulting probability distribution, leading us to give up on the non-trivial the assumption of accelerated Hamiltonians. Instead, we employ the Iterative Phase Estimation Algorithm (IPEA) with LPS and without SPS (Fig.\ref{fig:circuits}(a)). Specifically, we select the irrational phase $\phi = 2\pi/\sqrt{3}$ and evaluate it up to nine binary digits of accuracy, necessitating approximately $3*2^8 \approx 800$ consecutive gate applications, including syndrome extraction. We compute the mean $\bar{\theta}$ and standard deviation $\sigma$ of the resulting circular probability distribution (due to the phase's periodicity).

By averaging a Gaussian-like distribution \cite{o2019quantum}, the error scaling of the mean is determined to be $1/\sqrt{n}$, where $n$ represents the number of post-selected trials. The total measurement error is defined as the maximum between the digital error $2^{-m}$ (with $m$ denoting the desired precision) and the statistical error approximated in the limit of large $n$ by $\sigma/\sqrt{N(1-l_i)}$, where $N$ corresponds to the total number of algorithm trials. We define the minimal number of trials needed for the statistical error to reach to the digital error to be
\begin{equation} \label{eq:Nmin} N_{min} = 2^{2m}\frac{\sigma ^2}{1-l_i} \end{equation}

We anticipate the most significant improvement in the realistic case of a noisy sensor and perfect ancillas. Specifically, we initialize the sensor in the eigenstate $|+\rangle$ and measure the operator $R_x(\frac{2\pi}{\sqrt{3}})$ while assuming perfect ancillas and allowing only dephasing to occur to the sensor. The results, presented in Fig. \ref{fig:sensing_united}, demonstrate that even in the limit of infinite measurements, utilizing the LPS gadget enhances the accuracy of the mean compared to the ideal scenario (without noise). This improvement continues until the physical control method achieves the desired digital accuracy. Notably, Fig. \ref{fig:sensing_united}(c-d) reveals the existence of noise thresholds, indicating substantial improvements of up to an order of magnitude in error estimation and in the total number of experiments when comparing physical and logical control. Supplementary results can be found in \cite{Supplemental}.

In conclusion, we have introduced the concept of logical-physical qubit interaction. We have identified a parameter regime where its utilization is advantageous, considering circuit depth and worst case single gate fidelity, particularly in the presence of dephasing as the primary source of error \cite{chapeau2020fourier, chernyavskiy2019fidelity}. By encoding noise into a larger Hilbert space and employing post selection of purified states, we have demonstrated improvements of up to an order of magnitude in algorithmic quantum sensing within various real-world sensing scenarios. Logical post selection has proven to be effective in cases where ancilla qubits are more resilient to noise compared to the sensor or when Hamiltonian fast-forwarding is possible. The hybrid logical-physical interaction has considerable applications in further research: it can be efficient in algorithms that require long-lived ancilla qubits like state distillation, error mitigation and algorithmic sensing \cite{czarnik2021qubit, piveteau2021error, huggins2021virtual}. However, the drawback of information loss due to post selection should be noted, and recent studies are exploring methods for simulating without the need for post selection \cite{ippoliti2021postselection}. This concept of LPS opens up new frontiers, such as designing entangling gates between a logical qubit and a physical qubit with minimum error or error propagation, potentially leveraging fault-tolerant flag techniques\cite{chao2018fault, chao2020flag, reichardt2020fault,debroy2020extended}. Our work showcases the suitability of the five qubit code for enabling this type of interaction and we believe that it can also be implemented with today's most promising codes, such as the surface codes \cite{dennis2002topological}.
\par We acknowledge the Israeli Science Foundation (grants 963.19 and 2323.19) and Tuvia Gefen and Alex Retzker for fruitful discussions.

\nocite{*}
\bibliography{bibliography}

\pagebreak
\widetext
\begin{center}
\textbf{\large Supplemental Materials: Hybrid Logical-Physical Qubit Interaction as a Post Selection Oracle}
\end{center}

\newcommand{\beginsupplement}{%
        \setcounter{table}{0}
        \renewcommand{\thetable}{S\arabic{table}}%
        \setcounter{figure}{0}
        \renewcommand{\thefigure}{S\arabic{figure}}%
        \setcounter{equation}{0}
        \renewcommand{\theequation}{S\arabic{equation}}%
     }
     
\beginsupplement

\section{Theoretical Background} \label{appendix:TheoreticalBackground}
\subsection{Difference Measures between Density Matrices}
We utilize one of the standard definitions of the Distance between two quantum states $\rho, \sigma$ to be
\begin{equation} \label{eq:D}
    D(\rho,\sigma) = Tr|\rho-\sigma|^2
\end{equation}
With $|A|=\sqrt{A^\dagger A}$ the positive square root of $A^\dagger A$. It is straightforward to prove that our definition of \textit{Distance} behaves the same as the widely used \textit{Trace Distance} defined as $\frac{1}{2}Tr|\rho-\sigma|$ \cite{nielsen2002quantum}.

We also use the regular definition of \textit{Fidelity}, which is a good metric for state overlap, with the standard property of saturating to 1 when the states become identical.
\begin{equation} \label{eq:F}
    F(\rho,\sigma) = Tr\sqrt{\rho^{1/2}\sigma\rho^{1/2}}
\end{equation}

\subsection{Quantum Phase Estimation}
Quantum phase estimation is a family of algorithms. Although it is quite widely known in the quantum information community, for completeness we briefly overview the main ideas of the algorithm, and two iterative implementations of it. Suppose a unitary operator $U$ has an eigenvector $|u\rangle$ with eigenvalue $e^{2\pi i \phi}$, where the value of $\phi$ is unknown. The goal of the phase estimation algorithm is to estimate $\phi$. To perform the estimation, we assume we have available black boxes capable of preparing the state $|u\rangle$ and performing controlled-$U^{2^{j}}$ operations for some positive integer $j$. The algorithm uses two quantum registers, one for the measured operator $U$ and one for ancilla qubits needed for the computation. Phase estimation was first introduced by Kitaev \cite{kitaev1995quantum}.
\subsubsection{Kitaev's Iterative Phase Estimation}  
By introducing the $m$-bit approximation denoted as $\tilde{\phi} = 0. \phi _1 \phi _2 \cdots \phi _m$, and defining $\alpha _k =2^{k-1} \tilde{\phi}$, the utilization of the circuit illustrated in figure 1 (a) of the main text leads to the following relations: for the application of $K=I$, we obtain $\cos(2 \pi \alpha _k) = 2P_I(0|k)-1$, and for the application of $K=S$, we have $\sin(2 \pi \alpha _k) = 1-2P_S(0|k)$, with $P_{I(S)}$ the probability to measure $|0\rangle$ after applying $K=I(S)$ to measure the $k'th$ digit. This information is sufficient for the extraction of $\alpha _k$. Upon obtaining all $\alpha _k$ values for $k$ ranging from 1 to $m$, we can then retrieve the approximation $\tilde{\phi}$ using algorithm \ref{alg:Kitaev} \cite{svore2013faster}.

\begin{algorithm}
\SetAlgoLined
\KwResult{$\tilde{\phi}=0.\phi _1 \phi _2 ... \phi _{m+2}$ the ($m$+2)-bit approximation to the phase $\phi$}
Estimate all $\alpha _k$ using the circuit in figure 1 (a) of the main text\;
Set $0.\phi_m  \phi_{m+1} \phi_{m+2} = \beta _m$ where $\beta _m$ is the closest octant $\{\frac{0}{8},\frac{1}{8}, ... , \frac{7}{8}\}$ to $\alpha _m$\;
  \For{$j=m-1$ \KwTo $1$}{
  $\phi_j = \left\{
	\begin{array}{ll}
		0  & \mbox{if } |0.0\phi_{j+1}\phi_{j+2}-\alpha_j| _{mod 1} < 1/4 \\
		1  & \mbox{if } |0.1\phi_{j+1}\phi_{j+2}-\alpha_j| _{mod 1} < 1/4
	\end{array}
\right.$  }
return $\tilde{\phi} = 0.\phi_1\phi_2...\phi_{m+2}$
 \caption{Kitaev Estimator}
 \label{alg:Kitaev}
\end{algorithm}

\subsubsection{Iterative Phase Estimation Algorithm}
The Iterative Phase Estimation (IPEA) method, as depicted in figure 1 (a) of the main text \cite{dobvsivcek2007arbitrary}, utilizes a single ancilla qubit for performing phase estimation. This characteristic renders it particularly valuable in the context of Noisy Intermediate Scale Quantum (NISQ) computers, given the current limitations on the simultaneous utilization of multiple qubits. It is worth noting that the implementation of this circuit is challenging, and recent research \cite{corcoles2021exploiting} has demonstrated its feasibility. Several recent studies \cite{daskin2014universal, johnstun2021optimizing} have highlighted the significant potential of the IPEA algorithm. A notable advantage of this algorithm is its success probability, which, in an ideal scenario, remains independent of the desired number of measured digits.

\section{The 5-Qubit Code} \label{appendix:Code}
\begin{figure*} 
  \centering
  \includegraphics[width=1.0\textwidth]{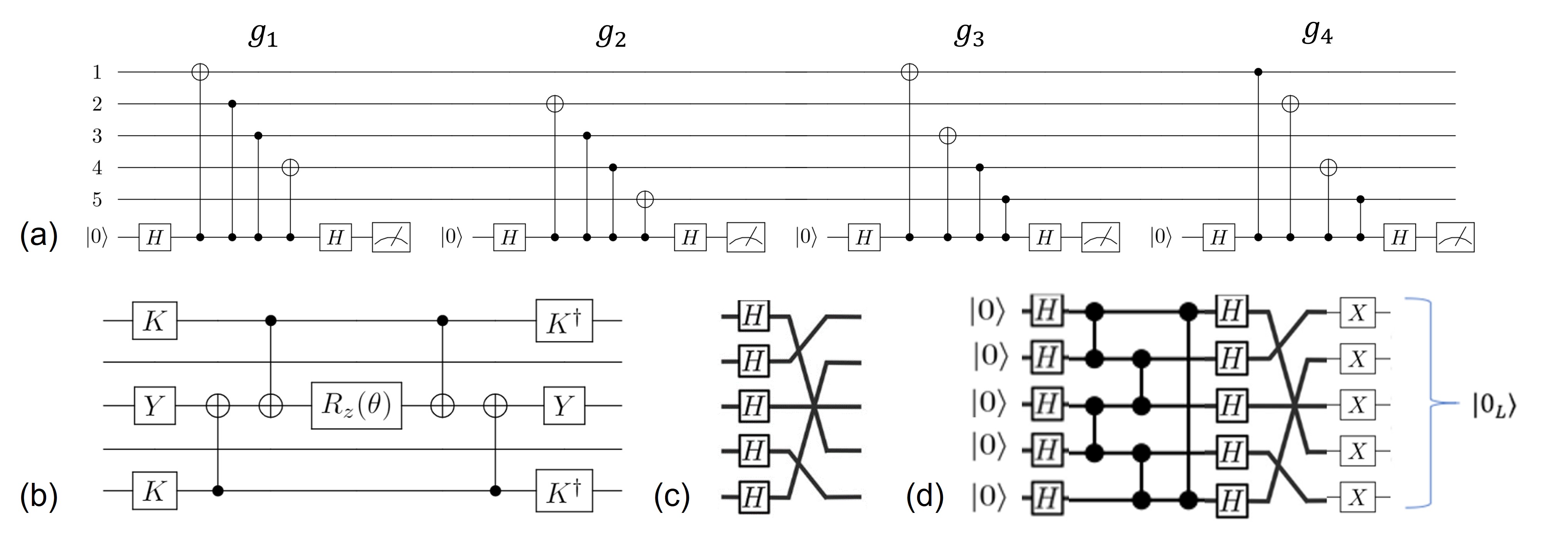}
  \caption{(a) Logical Post Selection is essentially applying syndrome extraction and post selecting the results that are in the code (trivial syndrome). (b) The logical $R_z(\theta)$ by \cite{yoder2016universal}. (c) The logical Hadamard gate by \cite{yoder2016universal}. (d) State preparation for the 5-qubit code, inspired by \cite{chao2018fault}.}
  \label{fig:appendix_circuits}
\end{figure*}

The 5-qubit code \cite{laflamme1996perfect} is a stabilizer code defined by the stabilizers in table \ref{table:generators} or by the logical basis states defined in equations \ref{eq:BasisState1}, \ref{eq:BasisState0}. As mentioned in the main text, any error in one or two qubits will result in measuring a non trivial syndrome. This phenomena of the 5-qubit code is presented in table \ref{table:causes}. It's basis states are the following:
\begin{equation} \label{eq:BasisState1}
\begin{split}
    |1\rangle_L = \frac{1}{4} \bigr[ |11111\rangle + |01101\rangle + |10110\rangle + 01011\rangle\\ 
+|10101\rangle - |00100\rangle - |11001\rangle - 00111\rangle\\-|00010\rangle - |11100\rangle - |00001\rangle - |10000\rangle\\-|01110\rangle - |10011\rangle - |01000\rangle + |11010\rangle \bigr]
\end{split}
\end{equation}

\begin{equation} \label{eq:BasisState0}
\begin{split}
|0\rangle_L = \frac{1}{4} \bigr[ |00000\rangle + |10010\rangle + |01001\rangle + 10100\rangle\\ 
+|01010\rangle - |11011\rangle - |00110\rangle - 11000\rangle\\-|11101\rangle - |00011\rangle - |11110\rangle - |01111\rangle\\-|10001\rangle - |01100\rangle - |10111\rangle + |00101\rangle \bigr]
\end{split}
\end{equation}
All necessary logical gates that were not defined in the main text are depicted in figure \ref{fig:appendix_circuits}.

\begin{table}
\begin{tabular}{ |p{3cm}|p{3cm}|  }
 \hline
 \multicolumn{2}{|c|}{ Generator Table for the 5-Qubit Code} \\
 \hline
 $g_1$ & $X_1Z_2Z_3X_4$ \\
 $g_2$ &   $X_2Z_3Z_4X_5$  \\
 $g_3$ &$X_1X_3Z_4Z_5$ \\
 $g_4$ &$Z_1X_2X_4Z_5$ \\ 
 \hline
\end{tabular}
\caption{Generator Table for the 5-Qubit Code}
\label{table:generators}
\end{table}

\begin{table}
\centering
\begin{tabular}{ |p{3cm}|p{8.5cm}|  }
 \hline
Error Syndrome & Possible Cause \\ 
\hline
0000&IIIII\\
\hline
0001&XIIII, IIYYI,  IIZIX,  IXIZI,  IYXII,  IZIIZ,  IIIXY\\
\hline
0010&IIIIX,  IIXYI,  IYYII,  IZIXI,  XIZII,  YXIII,  ZIIZI\\
\hline
0011&IIZII, IIIXZ,  IYIYI,  IZIIY,  XIIIX,  YIIZI,  ZXIII\\
\hline
0100&IIIXI, IIZIZ,  IXYII,  IZIIX,  XIIIY,  YYIII,  ZIXII\\
\hline
0101&IIIIY, IIYZI,  IXIYI,  IZZII,  XIIXI,  YIXII,  ZYIII\\
\hline
0110&IZIII, IIIXX,  IIXZI,  IIZIY,  XIIIZ,  YIYII,  ZIIYI\\
\hline
0111&IIIIZ, IIZXI,  IXXII,  IYIZI,  XZIII,  YIIYI,  ZIYII\\
\hline
1000&IIXII, IIIYX,  IXIIZ,  IZIZI,  XYIII,  YIIIY,  ZIIXI\\
\hline
1001&IYIII, IIIZZ,  IIYIX,  IIZYI,  XIXII,  YIIXI,  ZIIIY\\
\hline
1010&IIIYI,  IIXIX,  IXIIY,  IYZII,  XIYII, YIIIZ,  ZZIII\\
\hline
1011&IIYII, IIIZY,  IXIXI,  IYIIX,  XIIYI,  YZIII,  ZIIIZ\\
\hline
1100&ZIIII, IIIZX,  IIXXI,  IIYIZ,  IXZII,  IYIIY,  IZIYI\\
\hline
1101&YIIII, IIIYZ,  IIXIY,  IIZZI,  IXIIX,  IYIXI,  IZYII\\
\hline
1110&IIIZI,  IIYIY,  IYIIZ,  IZXII,  XXIII,  YIZII,  ZIIIX\\
\hline
1111&IXIII, IIIYY,  IIXIZ,  IIYXI,  XIIZI,  YIIIX,  ZIZII\\
\hline

\end{tabular}
\caption[Logical Post Selection - Error Probability Scaling]{Error syndrome and possible cause for the 5 qubit code. Causes with errors in more then 2 qubits are neglected. The error syndrome is $\langle g_1, g_2, g_3, g_4\rangle$.}
\label{table:causes}
\end{table}

\section{Scaling of the Error Probability} \label{appendix:ErrorScaling}
In this section we dive deeper into understanding the notion of logical post selection, and analyse the new scaling of the error probability. The incentive of developing a deeper understanding appears in table \ref{table:causes}, where we can see that errors in two or less qubits will result in measuring the trivial syndrome. In our error analysis we follow the explanation by Neilsen and Chuang \cite{nielsen2002quantum}.

Here, we present an example of a particular simple error analysis of LPS for the five qubit code. We assume the depolarising channel with probability $p$ acts on the state, giving
\begin{equation} \label{equation:depolarisingChannel}
    \rho \rightarrow (1-p)\rho + \frac{p}{3}(X\rho X + Y\rho Y + Z\rho Z)
\end{equation} 
For a simple one physical qubit case, taking an initial pure state $\rho=|\psi\rangle\langle\psi|$, we get process fidelity of
\begin{align*}
    F&=\sqrt{\langle\psi|\rho|\psi\rangle} \\
    &=\sqrt{(1-p) + \frac{p}{3}[\langle\psi|X|\psi\rangle^2 + \langle\psi|Y|\psi\rangle^2 + \langle\psi|Z|\psi\rangle^2]}
\end{align*}
This expression reaches the lowest fidelity for $|\psi\rangle=|0\rangle$, with:
\[F=\sqrt{1-\frac{2}{3}p} = 1-\frac{p}{3}+O(p^2)\]
Now, for the logical qubit. Assume we encode one qubit of information into $n$ physical qubits, each goes through a depolarizing channel $\varepsilon$ with probability $p$, as in equation \ref{equation:depolarisingChannel}. Then the channel's action on a state $\rho$ becomes:
\[\varepsilon^{\otimes n}(\rho) = (1-p)^n\rho + \sum_{j=1}^{n}\sum_{k=1}^{3} (1-p)^{n-1}\frac{p}{3}\sigma_k^j\rho\sigma_k^j\] \[ + \sum_{j_1=1}^{n}\sum_{j_2=j_1+1}^{n}\sum_{k_1=1}^{3}\sum_{k_2=1}^{3} (1-p)^{n-2}\frac{p^2}{9} \sigma_{k_1}^{j_1}\sigma_{k_2}^{j_2}\rho\sigma_{k_2}^{j_2}\sigma_{k_1}^{j_1} + ... \]
With $\sigma_k^j$ being the $k$'th Pauli operator acting on the $j$'th qubit. The first element represents one faulty qubit and the second represents two faulty qubits, and the dots represent errors in more then 2 qubits, which are neglected. Now, after performing LPS, each element in this sum will be returned to the state $\rho$ given $\rho$ was in the code:
\begin{align*}
(R\otimes &\varepsilon^{\otimes n})(\rho) = \\ &\left[(1-p)^n + np(1-p)^{n-1} + {\binom{n}{2}}p^2(1-p)^{n-2}\right]\rho 
\end{align*}
And finally, the fidelity $F$ remains:
\begin{equation} \label{equation:scaling}
    \begin{split}
        F&\geq\sqrt{(1-p)^{n-2}(1+(n-2)p+(\frac{n(n-1)}{2}-n+1)p^2)} \\
&=1-\frac{1}{12}n(n^2-3n+2)p^3+O(p^4)
    \end{split}
\end{equation}
Giving a $p^3$ dependence and confirming our intuition from table \ref{table:causes}.

\begin{figure*}
  \centering
\includegraphics[width=0.9\textwidth]{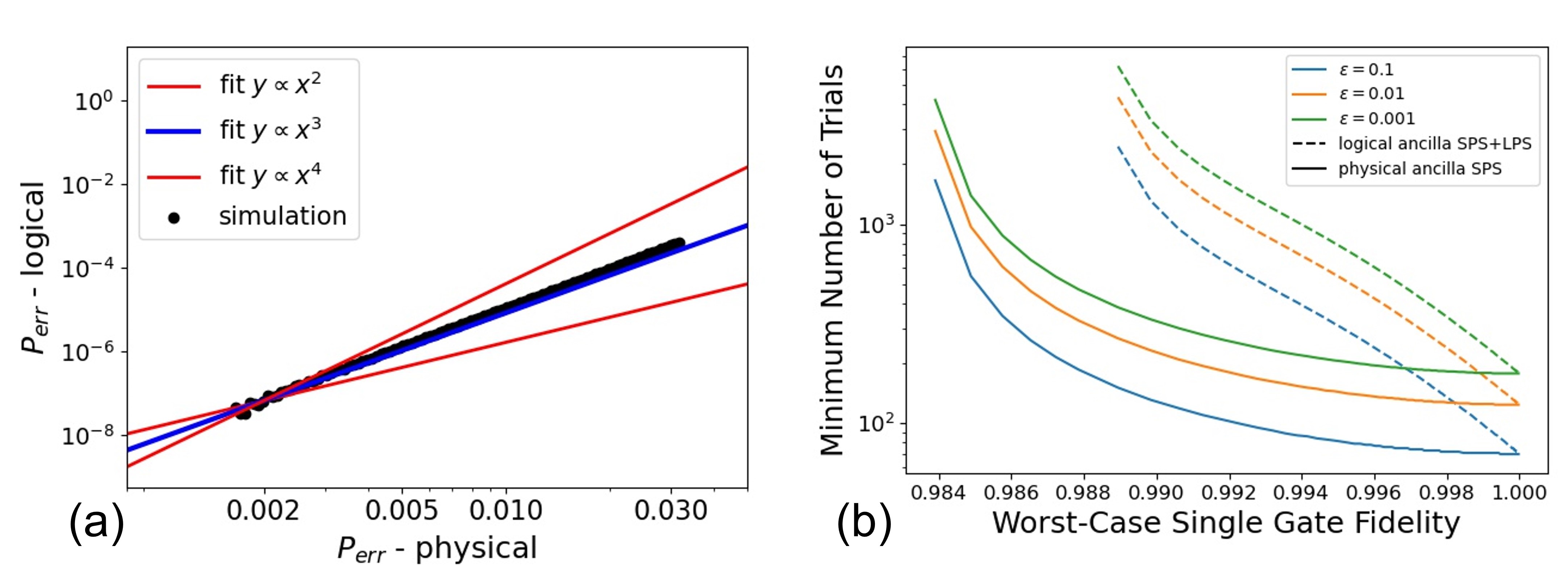}
  \caption{(a) Confirming the error probability scaling as predicted by equation \ref{equation:scaling}. Error probability extracted from fidelity according to equation \ref{equation:ErrorProb}. (b) Minimal number of trials for Kitaev's algorithm to succeed according to Eq. \ref{equation:MinimalNumberofTrials}, obtained by the data of figure 2 (c) of the main text.}
\label{fig:scalingError} \label{fig:KitaevCosineAll}
\end{figure*}

Implementing a depolarizing channel of the length of one gate and applying perfect (not noisy) syndrome extraction, we extract the error probability polynomial as in equation \ref{equation:ErrorProb}.
\begin{equation}\label{equation:ErrorProb}
P_{error} = 1-F
\end{equation}
Here, $F$ represents the fidelity calculated according to equation \ref{eq:F}. The analysis illustrated in figure \ref{fig:scalingError} (a) confirms that the best approximation to the scaling is not a second or fourth degree polynomial, but rather a third degree polynomial, as anticipated based on our theoretical derivations. We again emphasize that we assume no classical measurement errors. 

Transversality in the context of quantum error correction is an attribute of a quantum gate. A logical gate is transversal if in it's decomposition there is no entangling operation that allows the propagation of an error from one qubit to another, where both qubits are a part of a bigger, logical qubit. The implemented iterative versions of quantum phase estimation from the main text are not transversal, so the above analysis is not exact for them. However, a fault tolerant implementation is possible, by using flag fault-tolerance \cite{chao2018fault}.

\section{Additional Results - Kitaev QPE}
In this section we develop an understanding of the use of accelerated Hamiltonians in the resource-limited case. Here the settings are the same as in figure 2 (c) of the main text, i.e: perfect sensor and noisy ancilla. We define the asymptotic probability vector $\mathcal{P}$ to be comprised of the estimated probability to sample each result after in infinite number of samples. We define the finite sampled probability vector $\tilde{P}$ to be comprised of the probability to sample each result after in finite number of samples. $N$ is the total number of trials and $m=N(1-l_i)$ is the number of successful trials, with $l_i$ being the same lost information as in the main text. A probability distribution obtained from a noisy circuit has a subscript $P_n$ and a probability distribution obtained from an ideal, not-noisy circuit has a subscript $P_i$.
Expressing the distance between the vectors of an ideal probability and a noisy probability using equation \ref{eq:D} we get $|d|=|\mathcal{P}_n-\mathcal{P}_i|=D/\sqrt{2}$. The error in the estimated noisy probability $\tilde{P}_n$ is, by the addition rule for statistical and systematic errors, 
\begin{align*}
    \Delta \tilde{P}_n &= \sqrt{|\tilde{P}_n-\mathcal{P}_n|^2+|\mathcal{P}_n-\mathcal{P}_i|^2}\\
    &=\sqrt{|\tilde{P}_n-\mathcal{P}_n|^2+\frac{D^2}{2}}
\end{align*}
Here we aim to estimate the minimal number of trials required for the algorithm to succeed with probability larger than $1-\epsilon$, where $\epsilon$ defines our confidence in the result. To do this we closely follow \cite{ahmadi2010quantum}.

Following reference \cite{ahmadi2010quantum} we demand that the error $\Delta \tilde{P}_n$  be confined, such that the probability that the algorithm succeeds
is
\begin{equation*}
    Pr\left(\Delta \tilde{P}_n<\frac{2-\sqrt{2}}{4}\right)
\end{equation*}
This equals to
 \begin{equation*}Pr\left(|\tilde{P_n}-\mathcal{P}_n|^2<\left(\frac{2-\sqrt{2}}{4}\right)^2-\frac{D^2}{2}\right)
 \end{equation*}
 $D$, the distance between output states, and the lost information, $l_i$, depend on many features, one of which is the noise or (in this work) the dephasing time $T_2$. We denote these dependencies as $D(T_2), l_i(T_2)$ respectively. Demanding positivity of the right hand side results in a condition for the algorithm to succeed in any probability. The algorithm fails for noise $T_2$ that holds
  \begin{equation} \label{failCond}
     D(T_2) > \frac{\sqrt{2}-1}{2}
 \end{equation} 
We conclude that the probability for the algorithm to fail (by applying Chernoff bound) \cite{ahmadi2010quantum} is
\begin{equation*}
\begin{split}
     Pr\left(|\tilde{P_n}-\mathcal{P}_n|\geq\sqrt{\left(\frac{2-\sqrt{2}}{4}\right)^2-\frac{D^2}{2}}\right) \leq \\ 2e^{-2((\frac{2-\sqrt{2}}{4})^2-\frac{D^2}{2})m}
\end{split}
\end{equation*}
and we demand that the probability to succeed is
$$Pr\left(|\tilde{P_n}-\mathcal{P}_n|<\sqrt{\left(\frac{2-\sqrt{2}}{4}\right)^2-\frac{D^2}{2}}\right) \geq 1-\epsilon$$
This gives us a minimum of
\begin{equation} \label{equation:MinimalNumberofTrials}
    N>\frac{\ln(\frac{2}{\epsilon})}{2(1-l_i(T_2))((\frac{2-\sqrt{2}}{4})^2-\frac{(D(T_2))^2}{2})}
\end{equation}
trials for the algorithm to succeed with probability of success $p \geq 1-\epsilon$. 
Calculating this value for noise of different strengths and for a number of different $\epsilon$'s we get the expected result of Fig. \ref{fig:KitaevCosineAll} (b), showing no improvement of the logical approach over the physical one.

\section{Additional Results - IPEA}
\subsection{Sensing}
\begin{figure*}
    \centering
    \includegraphics[width = 0.9\textwidth]{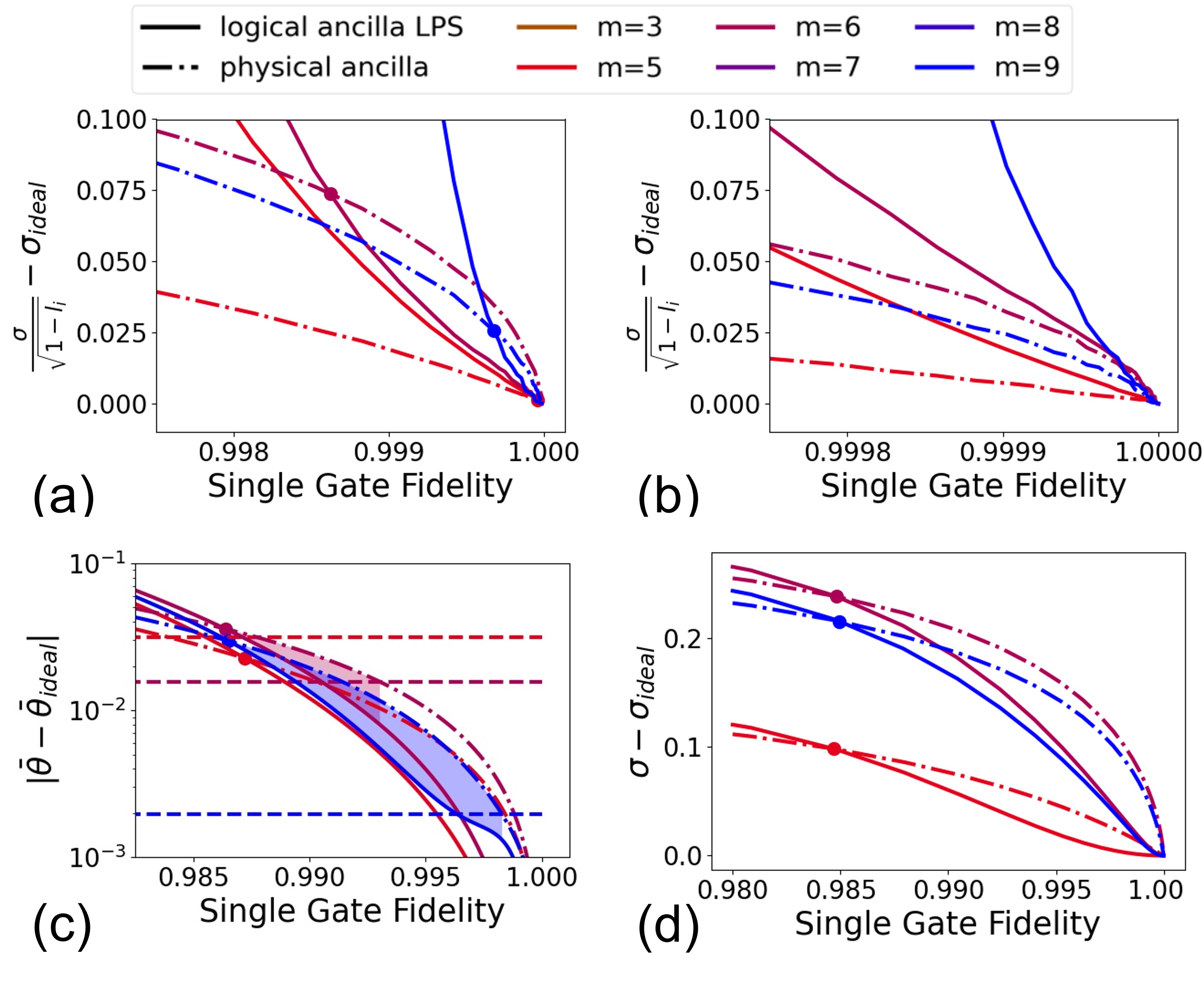}
    \caption{Additional results for the histogram of possible results of IPEA in the presence of noise. (a-b) Error in estimating the phase of non-accelerated signal Hamiltonian $R_z(\theta)$ with sensor initialized in the state $|1\rangle$, in comparison to the ideal, using IPEA. The relative error is plotted for desired precision of 5,6,9 digits, while continuous line represents logical ancilla and LPS, and dotted-dashed line represents a physical ancilla. (a) The ancillas are perfect and the sensor is noisy, susceptible only to amplitude damping. The relevant thresholds are the crossings of dashed and continuous lines of the same color. (b) The ancillas are noisy, susceptible only to dephasing, and the sensor is perfect.  It is clear that with non-accelerated Hamiltonian and perfect sensor, adding noisy ancillas does not improve sensing capabilities with IPEA. (c) Fig.3b from the paper, in a logarithmic scale. (d) Base standard deviation of the histogram. It is apparent that the thresholds in those cases, which does not take lost information into account, are around 0.985 worst-case single qubit gate fidelity.}
    \label{fig:IPEA_supp} \label{fig:SadditionalIPEA}
\end{figure*}

Here we present the two additional scenarios complementing the one in the main text: 
\begin{itemize}
    \item \textit{Fig. \ref{fig:IPEA_supp} (a)} - The sensor is put in the excited state and is susceptible to $T_1$ noise, while the ancilas are perfect, measuring $R_z(\frac{2\pi}{\sqrt{3}})$. In this scenario thresholds are observed similarly to the scenario studied in the main text.
    \item \textit{Fig. \ref{fig:IPEA_supp} (b)} - The sensor is perfect (not noisy) and is put in the excited state while the ancillas are noisy, subject to dephasing, measuring $R_z(\frac{2\pi}{\sqrt{3}})$. In this scenario we observe no crossing points, which means that adding noisy ancillas to a perfect sensor does not improve sensing capabilities in the sense of sensing time.
\end{itemize}
To complete the picture provided by Fig.2d, Fig3 in the main text, we present some additional results on the structure of the histogram of possible results. In figure \ref{fig:SadditionalIPEA} (c) is Fig.3b of the main text in a logarithmic scale, and  \ref{fig:SadditionalIPEA} (d) presents the bare standard deviation of the histogram.

\subsection{Computing}
In some cases we are instead interested in applying the algorithm only once and assess it's probability to give the right result. In such cases, where the sensing time or computation time has a lower importance, it is convenient to have an oracle stating whether the result is reliable or not. Implementation of IPEA is possible in two types of systems - those that enable a large amount of parallelism and those that do not. We call these systems ensemble systems, and superconducting systems, respectively. Here we first explain how to calculate an upper bound to the error probability of the ensemble scenario, and then we show our simulated error probability for the superconducting scenario.
In both cases the success probability is defined to be:
\begin{equation} \label{eq:Perr}
    P_{success} = \sum_{\phi \in [\tilde{\theta},\tilde{\theta}+2^{-m}]} p(\phi)
\end{equation}
Where $\theta$ is the measured phase, $\tilde{\theta}$ is the best $m-$bit approximation to it from below $\phi$ is a possible result of the algorithm, $p(\phi)$ is the probability to measure $\phi$ and $P_{error}=1-P_{success}$.
\begin{enumerate}
    \item Ensemble systems,  where it is convenient to apply the whole algorithm in parallel on a large number of qubits and measure result statistics. Assume an experiment contains $n$ runs of the whole algorithm, and the result is taken to be a majority vote on the most probable interval of phases. Then
\begin{align*}
    P_{error} &\leq P(\text{less then $n/2$ trials in the interval}) \\  &= P(\text{0 successes in $n$ trials}) +P(\text{1 successes in $n$ trials}) \\ &+\cdots+P(\text{$\lfloor n/2\rfloor$ successes in $n$ trials})
\end{align*}
Each of the probabilities in the above formula is distributed binomialy, and so it is possible to exactly compute an upper bound for $P_{error}$ \cite{butler1993distribution}.
    \item Superconducting systems, where it is convenient to measure the digit in each iteration a number of times and take a majority vote. Assume the experiment contains $n$ runs for each bit and the result for the bit is taken to be a majority vote. This has the largest impact only on the first (least significant) digits. There are only two possible results - 0 is measured or 1 is measured. Thus the probaility to measure 0 is updated in our simultaion to be the probability to measure 1 less then half of the times, and vice versa:
\begin{align} \label{eq:SPupdate}
        p(0) \leftarrow \sum_{k=0}^{\lfloor n/2 \rfloor} \binom nk p_1^k p_0^{n-k}  &&
    p(1) \leftarrow \sum_{k=0}^{\lfloor n/2 \rfloor} \binom nk p_0^k p_1^{n-k}
\end{align}
And the error probability is defined as in Eq.\ref{eq:Perr}.
\end{enumerate}

 \begin{figure*}
    \centering
    \includegraphics[width = 1.0\textwidth]{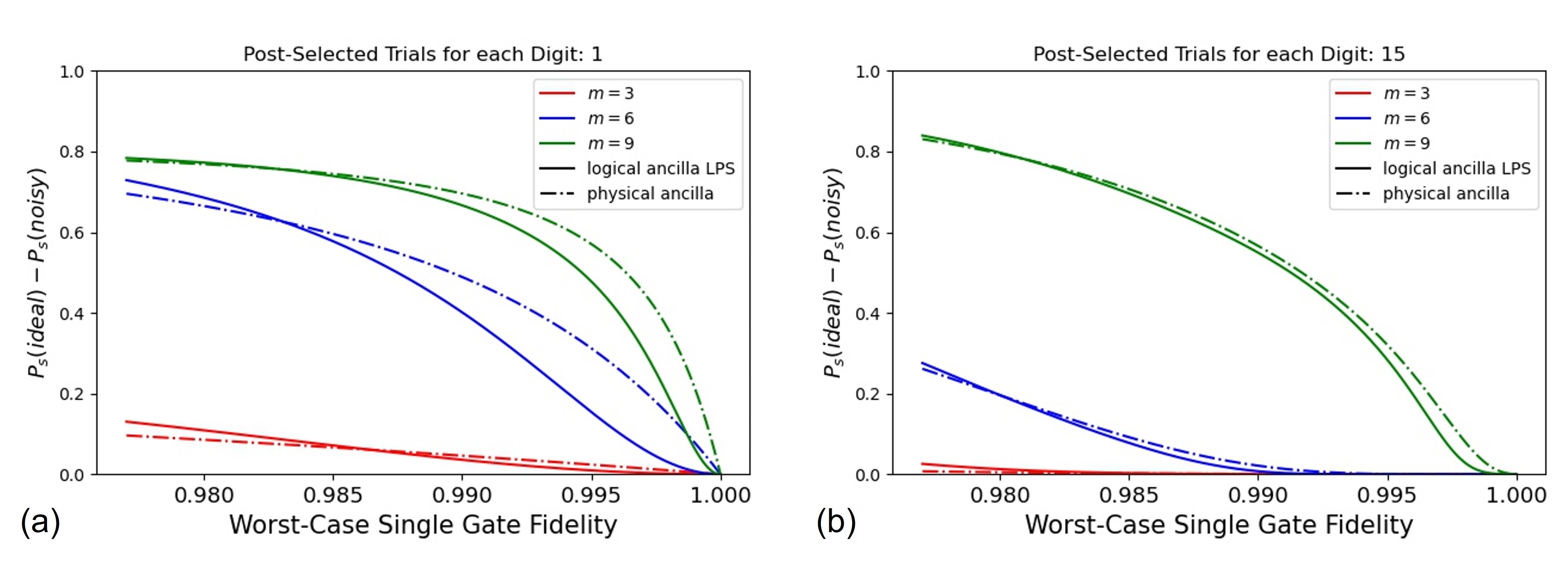}
    \caption{Scaling of the success probability as a function of noise for different $n$'s. The histogram is calculated by Eq.\ref{eq:SPupdate} and the success probability is calculated by \ref{eq:Perr} under the assumption that the interval contains only two bins. It is apparent that increasing $n$ leads to some flattening of the probability difference and to a larger proximity of the physical and logical control methods.}
    \label{fig:Perr1dunited}
\end{figure*}

\begin{figure*}
    \centering
    \includegraphics[width = 1.0\textwidth]{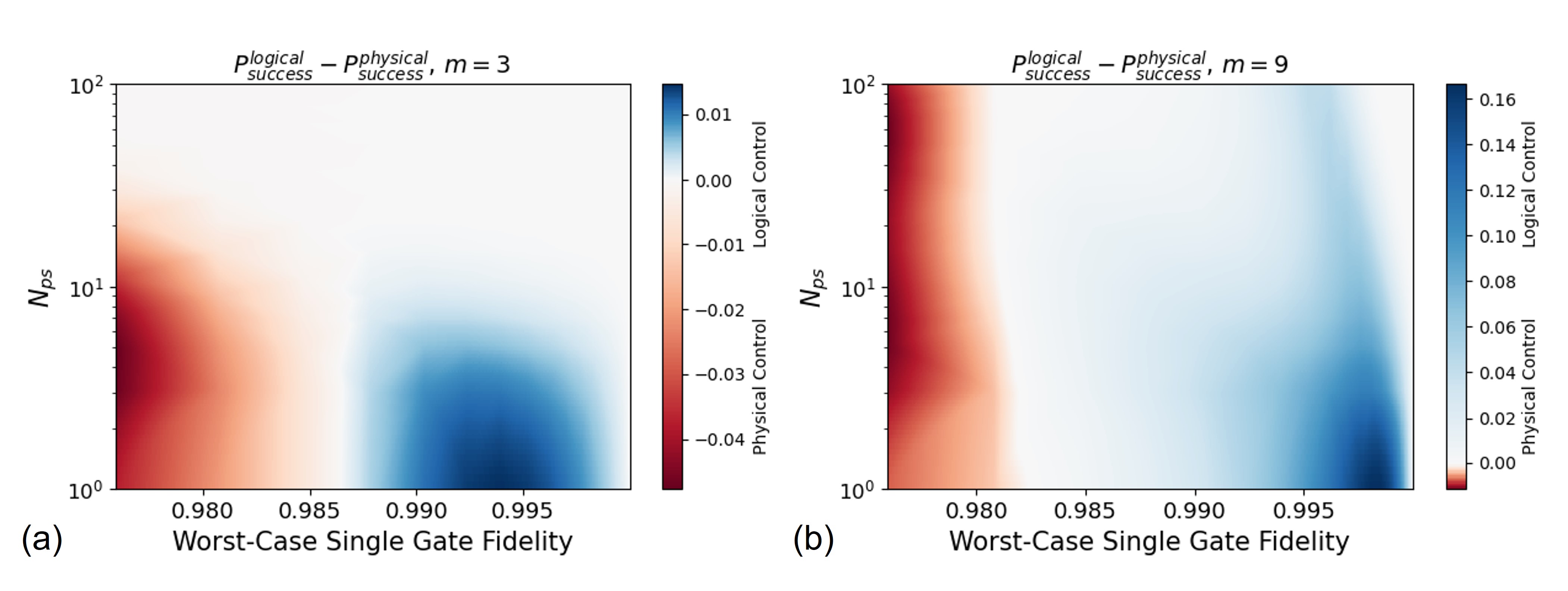}
    \caption{A two dimensional map of the success probability difference between physical and logical control methods for different desired precision of $m=3$ and $m=9$ digits. the X-axis is the worst case single qubit gate fidelity and the Y-axis is the number of post selected trials for each digit. It is apparent that from a circuit depth of around a few dozens of gates, there is a threshold in approximately 0.98-0.985 worst case gate fidelity. These results are well within the capabilities of today's hardware.}
    \label{fig:Perr2dunited}
\end{figure*}

\section{Simulation Details}
A detailed description of the simulation and a code guide are available in the following GitHub repository. Here we give only a brief description of the simulation.
\subsection{GitHub Code}
All relevant code is open for use in the following url: \url{https://github.com/nadavcarmel40/paper_recalc}. 
To install and use the package, just install \href{https://qutip.org/docs/latest/installation.html}{QuTiP} (\url{https://qutip.org/docs/latest/installation.html}) by following the instructions on the above web page.
The basic tool enabling the simulations can be found under the 'simulators' folder in the attached GitHub repository. 'BigStepSimulator' is a state-vector simulator and 'SmallStepSimulator' is a density matrix simulator. The state vector simulator is fast and can simulate quantum circuits without noise, and the density matrix simulator is slower and can simulate noisy circuits.

\subsection{General Description}
In this work, we use a full density matrix simulation, similar to Ref.\cite{cheng2021simulating}. We save the quantum state of $n$ qubit register as a 2-d matrix of dimensions $[2^n,2^n]$. Operators are saved as a $[2^n,2^n]$ matrix, and if the quantum state was initially $\rho$ then performing the operation $U$ on the density matrix is equivalent to updating the density matrix $\rho \rightarrow U\rho U^\dagger$.
 
\begin{figure}
    \centering
    \includegraphics[width = 0.9\textwidth]{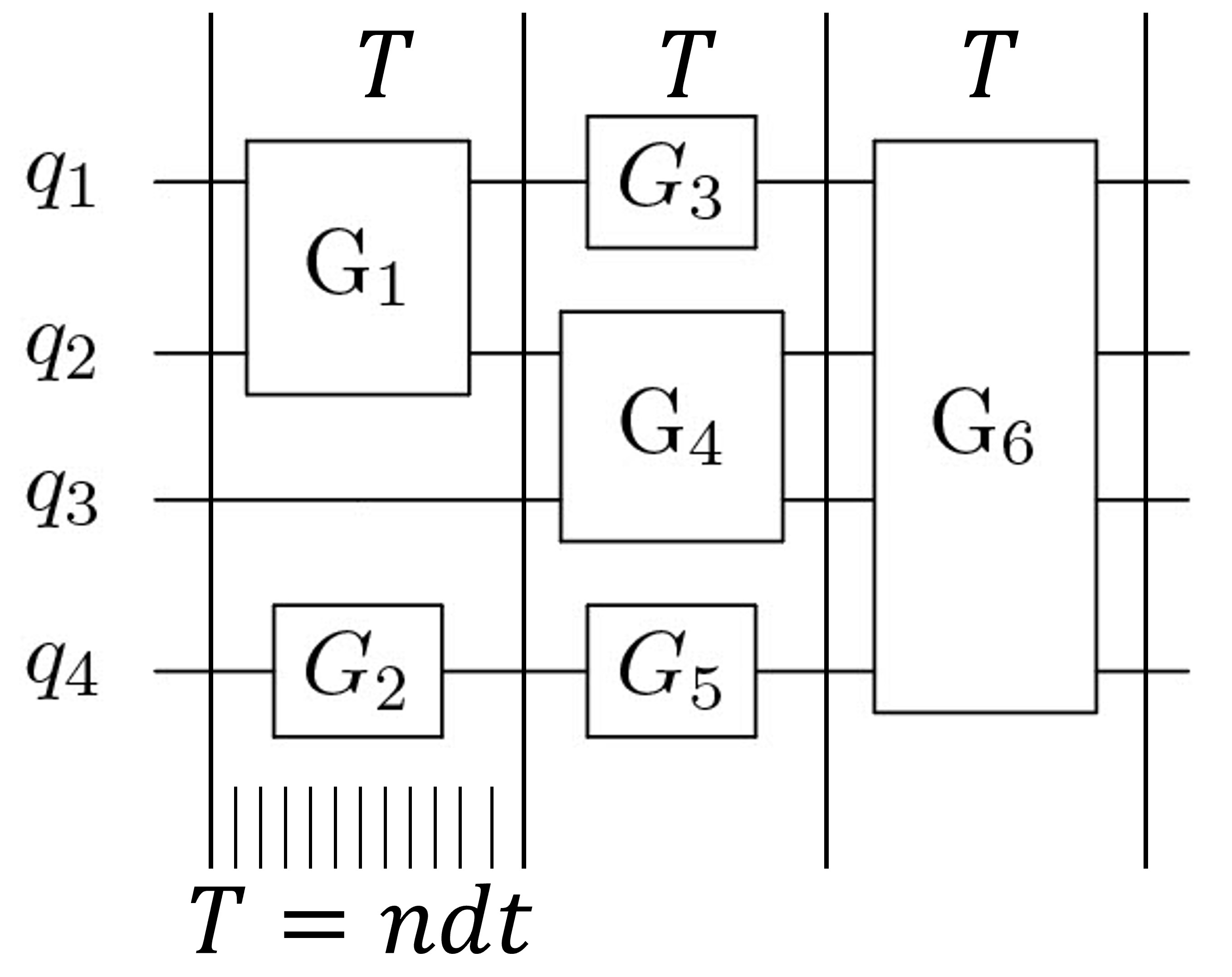}
    \caption[General Circuit Simulation]{Simulation of a general circuit. Each gate-step $T$ has length $T_{g}$ and is made up of $n=20$ small time-steps of size $T_{g}/n$. The simulation is made up of many small time steps, with repeated application of Krauss based decoherence on each qubit in one small time-step and gate-based evolution in the next small time-step.}
    \label{fig:generalSim}
\end{figure}

The noise in the simulation is based on Krauss operators. The simulation is thus made up of many small time steps, with repeated application of Krauss based decoherence on each qubit and gate-based evolution in each small time-step, see figure \ref{fig:generalSim}.

Defining the Pauli operator $\sigma_i^q$ acting on qubit $q$ as a tensor product of $\sigma_i$ in the $q$ index and Identity operators in all other indexes, the base Hamiltonian $H_0=\bigotimes_{q=1}^N \frac{\hbar \omega _{01}}{2}\sigma_z^q$ represents the free evolution of the quantum register, with $\omega_{01} = 2 \pi \cdot 6 [GHz]$. In our research we work in the rotating frame, but simulation is also possible outside of the rotating frame with our code package.

In each gate-step, possibly many gates act upon the register. Thus, we start with the base Hamiltonian $H=H_0$ and for each gate $G$ in the gate-step we find it's corresponding Hamiltonian $H_G$ given by $G=e^{iH_G}$ and update the Hamiltonian to be $H \rightarrow H+H_G$.
Now, we define the evolution operator $U$ to be $e^{iH\frac{dt}{T_{g}}}$, and we apply this evolution as described earlier for a total of $T_{g}/dt$ times with a decoherence step between each application of $U$.

 The main parameters used in each simulation are the number of qubits $N$, the time $T_{g}=n\cdot dt$ ($n=20$) of each gate-step, the dephasing time of qubit $q$, $T^q_2$ and the energy relaxation time $T^q_1$ of the same qubit. From these parameters we define the error rates for each process and qubit: $$p_{decay}^q=1-e^{-\frac{dt}{T^q_1}}$$ and $$p_{dephase}^q=1-e^{-\frac{dt}{T^q_2}}$$
The decoherence is then enacted upon the register through a for-loop on each qubit, updating the register state to be:

\[\rho = E_1^q \rho (E_1^q)^\dagger + E_2^q \rho (E_2^q)^\dagger\]
\[\rho = \left(1-\frac{P_{dephase}^q}{2}\right)\rho + \frac{P_{dephase}^q}{2}\sigma_Z^q \rho \sigma_Z^q\]

where the first equation uses the Pauli-matrix representation of the Krauss operators: $E_1^q = \frac{\sqrt{1-P_{decay}^q}}{2}(I-\sigma_Z^q)+\frac{1}{2}(I+\sigma_Z^q)$ and $E_2^q = \frac{\sqrt{P_{decay}^q}}{2} (\sigma_X^q+i\sigma_Y^q)$, and the second equation is a result of applying the phase damping channel's Krauss operators $\sum_i K_i \rho K_i^\dagger$ with probability $P_{dephase}^q$. These Krauss operators are given in the literature in their matrix form \cite{nielsen2002quantum}.

\subsection{Measurement and Lost Information}
 There are two kinds of measurements we perform - post selection measurements and probabilistic measurements. Here, we first refer to the post selection measurements. 
 Post selection measurements are done on the sensor qubit for SPS, on flag qubits for fault-tolerance, and on an additional ancilla qubit for LPS. To perform post selection measurements, we collapse the register state as defined below according to the preferred measurement outcomes. One could say we choose the system's trajectory.
 First, to add qubits to the simulation, we expand the register state with a tensor product to the additional qubit sub-spaces. Next, we perform the entangling operations with the additional qubits, and finally project on the trivial flag state $|0..0\rangle$ using the operator defined below. 
 For probabilistic measurements (e.g. Error correction measurements), to decide measurement outcome on the qubit group $A=\{q_{k_1},...,q_{k_n}\}$, remaining with $B=\{q_1,...,q_N\}/A$, we trace out B to get $\rho^A=Tr_B(\rho)$. Then, we define P as the diagonal of $\rho^A$ and $P'$ as the cumulative sum of P. we take a random number $0<x<1$ and find the first index $i$ such that $x<P'[i]$. The result of the measurement is the binary string of $i-1$.
 To collapse the quantum state to a state after measuring qubits in the group A, we use the following projector:
\[
  P_m = \bigotimes_{q=1}^N
  \begin{cases}
    I_2 & \text{if $q \notin A$} \\
    \frac{I+\sigma_Z^q}{2} & \text{if $q \in A$ and measurement result is $|0\rangle$} \\
    \frac{I-\sigma_Z^q}{2} & \text{if $q \in A$ and measurement result is $|1\rangle$}
  \end{cases}
\]
And after this projection operation $\rho \rightarrow P_m\rho P_m^\dagger$ we trace out the additional qubits.
 The procedure described above can cause numeric errors when the state decoheres for a long time, because the projection operation as described is not trace preserving. To have a valid density matrix, for each projection, say the $k$'th projection, we first save the state's trace as $Ps_k$ and then normalize the state. To calculate the portion of information that have been lost due to post selection, we use the following reasoning:
\begin{itemize}
    \item After one projection, we have lost $1-Ps_1$ information and remain with a state with trace $Ps_1$.
    \item After the second projection, we have lost  $Ps_1\cdot (1-Ps_2)$ more information.
    \item After the third projection, we have lost  $Ps_1\cdot Ps_2 \cdot (1-Ps_3)$ more information.
    \item After the $k$'th projection, we have lost  $Ps_1\cdot...\cdot Ps_{k-1} \cdot (1-Ps_k)$ more information.
\end{itemize}
Overall, this is the amount of \textit{lost information}:
\begin{equation} \label{equation:lostinformation}
    l_i = \sum_i ((\prod_k^{i-1} Ps_k)\cdot (1-Ps_i))
\end{equation}

\end{document}